\RequirePackage{fix-cm}

\documentclass[smallextended]{svjour3}       
\smartqed  

\usepackage{color, soul}

\usepackage{textcomp}
\usepackage{longtable}
\usepackage{lineno, natbib}
\usepackage{hyperref}
\usepackage{mathtools}

\usepackage{times}
\usepackage{helvet}
\usepackage{courier}
\usepackage{amsmath}
\usepackage{enumerate}
\usepackage{graphicx}
\usepackage[colorinlistoftodos]{todonotes}

\usepackage{url}
\newcommand{\leaveout}[1]{}

\begin{document}

\title{Empirically Grounded Agent-Based Models of Innovation Diffusion: A Critical Review
}


\author{Haifeng Zhang \and Yevgeniy Vorobeychik
}


\institute{Haifeng Zhang \and Yevgeniy Vorobeychik\at
              Electrical Engineering and Computer Science\\
              Vanderbilt University\\
              Nashville, TN, USA\\
              \email{\{haifeng.zhang, yevgeniy.vorobeychik\}@vanderbilt.edu}           
}

\date{Received: date / Accepted: date}

\maketitle

\begin{abstract}
Innovation diffusion has been studied extensively in a variety of
disciplines, including sociology, economics, marketing, ecology, and computer science. 
Traditional literature on innovation diffusion has been dominated by
models of aggregate behavior and trends.
However, the agent-based modeling (ABM) paradigm is gaining
popularity as it captures agent heterogeneity and enables fine-grained
modeling of interactions mediated by social and geographic networks. 
While most ABM work on innovation diffusion is theoretical,
empirically grounded models are increasingly important, particularly
in guiding policy decisions.
We present a critical review of 
empirically grounded agent-based models of innovation diffusion, developing a categorization
of this research based on types of agent models as well as
applications.
By connecting the modeling methodologies in the fields of information
and innovation diffusion, we suggest that the maximum
likelihood estimation framework widely used in the former is a
promising paradigm for calibration of agent-based models for innovation diffusion. 
Although many advances have been made to
standardize ABM methodology, we identify four major
issues in model calibration and validation, and suggest potential solutions.

\end{abstract}

\section{Introduction}

\subsection{Innovation Diffusion: Theoretical Foundations}

\emph{Diffusion} refers to the process by which an innovation is
adopted over time by members of a social system~\citep{rogers10,valente05}. 
Commonly, an \emph{innovation} refers to a new technology, but the
conceptual notion can be applied far more broadly to consider the
spread of ideas and practices.
\cite{rogers10} laid down the theoretical foundations of innovation
diffusion in his book, \emph{Diffusion of Innovations}, in which he
synthesizes studies in anthropology, sociology, and education, and
proposes a generic theory to explain the diffusion of innovations among
individuals and organizations.  
He suggests five characteristics of innovation to determine the rate of adoption:
\emph{relative advantage}, \emph{compatibility}, \emph{complexity}, \emph{trialability}, and
\emph{observability}. 
Rogers models human decision about adoption of innovation as a
multi-stage process, involving five stages: \emph{knowledge}, \emph{persuasion}, \emph{decision}, \emph{implementation}, and \emph{confirmation}.
Furthermore, he classifies individuals into five adopter
categories: \emph{innovators}, \emph{early adopters}, \emph{early majority},
\emph{late majority}, and \emph{laggards}. 
In addition to these high-level considerations, much
attention has been on the significance of social relationships and
influence in innovation diffusion (in contrast with, or complementary to, economic
considerations).
Starting with early groundwork~\cite{ryan43}, there has now been
extensive research on how social network structure, group norm, opinion leadership, weak ties, and critical mass impact diffusion of innovations~\citep{vale95,vale95b}.



\subsection{Mathematical Models of Innovation Diffusion}
Traditional mathematical models of innovation diffusion aim to model
aggregate trends, rather than individual decisions.
Numerous such models follow the framework of Bass model, which is one of the most influential models in marketing~\citep{Bass69,hopp04}. 
The Bass model was originally designed for forecasting sales of new consumer durables. 
The model assumes that the probability of adopting a product, given
the person has not yet adopted, is linearly related to the number of past adopters. 
The Bass model can be calibrated with aggregate sales data, and Bass
showed that it can qualitatively capture the S-shaped pattern of
aggregate adoption over time~\citep{pere10}. 


The Bass model has a number of limitations. First, it does not capture individual interactions. Indeed, the model explicitly assumes a fully connected and homogeneous network.
For innovation diffusion, this is an important drawback, as individual interdependence and communications are among the most significant aspects to understand innovation diffusion~\citep{valente05,rogers10}. 
The second criticism of the Bass model is that it does not include any decision variables that are of interest from a managerial perspective. 
The issue has been addressed later by incorporating the marketing mix variables, price, and advertising, into the diffusion model. For an extensive review of research in this direction, we refer readers to~\cite{mahajan00,meade06}. 
Nevertheless, these marketing mix variables are mostly designated for the entire market without a consideration of individual heterogeneity. 
Lastly, the predictability of the Bass model is often questioned. 
For example,~\cite{chan07} argue, that the model needs considerable
data around the critical point at which diffusion accelerates to be
effective, but once such data is available the value of the
Bass model becomes limited.


\subsection{Agent-Based Modeling for Innovation Diffusion}
Agent-based modeling (ABM) has emerged as another natural approach to
study innovation diffusion. 
Agent-based models are typically simulation models that capture
dynamic interactions among a (usually large) collection of individuals.
They were originally developed as a tool for complexity theory research~\citep{lewin99, holland95}, and have gained popularity in many scientific areas for the past decade~\citep{gilbert05, macal10, garcia11, macal16}.
The ABM paradigm offers two advantages for the study of innovation
diffusion: 
first, it facilitates the modeling of agent heterogeneity, and second,
it enables fine-grained modeling of interactions mediated by social networks.
Indeed, agent-based modeling has been applied in study of innovation
diffusion to aid intuition, theoretical exploration, and to provide policy
decision support~\citep{kiesling12}.


Traditional agent-based models are largely conceptual~\citep{axelrod97,epstein99}.
This use of ABMs as primarily conceptual tools is partly because they
are commonly considered as ideal~\emph{learning tools} for scientists
to understand a system under a variety of conditions by simulating the
interactions among agents. 
As a consequence, the simplicity of agent rules is commonly a crucial
consideration in the design of agent-based models.
Such simplicity, however, has given rise to criticism of the ABM
methodology as being ``toy models'' that do not reflect reality~\citep{garcia11}.
Moreover, an increasingly important criticism is that if ABMs are used
in any policy decision support, the predictive validity of the model
becomes paramount, and models that are primarily conceptual may be
inadequate for such tasks.

It is this increasing use of agent-based modeling to obtain policy
guidance that has motivated increasing use of empirically grounded
agent-based models.
Empirical agent-based models have recently experienced significant growth~\citep{kiesling12}. 
In these studies, empirical data are used to initialize simulation,
parameterize agent-based models, or to evaluate model validity. 
The explosion of high-resolution data sets, coupled with advances in
data analytics and machine learning have given rise to increased opportunities for
empirically grounding agent-based models, and this trend is likely to continue.
Our goal is to provide an overview of these empirically grounded
agent-based models developed with the goal of studying innovation diffusion.
Through a careful examination of these studies, we also aim to
identify potential methodological issues that arise, and suggest ways
to address these.

\subsection{Contributions}

The diffusion of new products has been an important topic for decades~\citep{mahajan90, mahajan00, meade06, chan07, pere10}.    
The prevalence of ABM approach can be glimpsed from a number of review
papers from disciplines like sociology~\citep{macy00},
ecology~\citep{matthews07}, and marketing~\citep{garcia05, hauser06, negahban14}.
For example,~\cite{garcia05} describes potential uses of ABM in market research
associated with innovations, exploring benefits and challenges of
modeling complex dynamical systems in this fashion.
~\cite{dawid06} surveys agent-based models of innovation diffusion
within a computational economics context.
~\cite{pere10} review diffusion models in the context of a
single market and cross-markets and brands.
To the best of our knowledge, the closest work to ours is a review of
agent-based simulations of innovation diffusion by~\cite{kiesling12},
who survey both theoretical and empirical work. 
In comparison with these past reviews, we make the following novel contributions:
\begin{enumerate}
\item We provide \emph{systematic} review of the empirical agent-based
  models of innovation diffusion. This is in contrast to the narrative
  review of the applied work as provided in~\cite{kiesling12}.
In particular, we offer a novel classification of agent adoption models as employed in the reviewed papers.
By highlighting the adoption models and their parameterization methods, we aim to bridge methodological gaps among domains and applications.    
We identified the papers to include in a rigorous and systematic manner.
In terms of scope, any work presenting an agent-based model using empirical data to simulate the diffusion of innovations was included. 
Our selection process combined results from multiple databases,
including Google Scholar and ScienceDirect, with extensive search for
relevant keywords, and back-tracking and forward-tracking reference
lists, while carefully screening out non-candidates.

\item Our review is \emph{comprehensive} and \emph{updated}. 
The collection of reviewed papers spans a \emph{superset} of the applications as covered in~\cite{kiesling12} and, indeed, a number of
significant efforts have emerged after 2012.
Notably, we also include a selection of papers from the literature on
information diffusion, a fast-growing area. These models rely on
principled machine learning techniques for model calibration based on
empirical observations of diffusion traces.
In addition, we exclude two (out of 15) papers from~\cite{kiesling12} which
are not empirically grounded.
In the end, we reviewed 43 papers, of which 30 (23 from years after 2011) were not included by~\cite{kiesling12}.

\item We provide a \emph{critical} review, assessing strength and
  weaknesses of the surveyed research.  
  Almost all surveyed papers followed standard modeling steps and presented their results systematically.   
  However, we conclude that the
  current literature commonly exhibits several major shortcomings in
  model calibration and validation\footnote{The concepts of
    calibration and validation are explained in Section~\ref{sse:cv} below.}.     
Addressing these issues would significantly increase the credibility of agent-based models. 
We, therefore, devote a section to an overview of existing validation
methods in the literature and an in-depth discussion of these issues and potential solutions. 

\end{enumerate}

\section{Categorization of Empirically Grounded ABMs of Innovation Diffusion}
\label{sec: emp_abms}
We review the burst of recent developments of empirically grounded agent-based models, which are examined through two dimensions: models and applications. 
First, to facilitate methodological comparison, we group the papers
into six categories which represent the specific approaches taken to
model individual agent decision processes: \emph{mathematical optimization based models}, \emph{economic models}, \emph{cognitive agent models}, \emph{heuristic models}, \emph{statistics-based models} and \emph{social influence models}.
Second, as we observe that modeling efforts span several domains, the next section offers an application-focused categorization.

The categorization in this section is aimed at qualitatively clustering the existing agent-based models with respect to their \emph{modeling methods}, which can be further characterized from several dimensions, such as behavioral assumption, data granularity, internal structure, calibration and validation. 
The six categories we identified present a comprehensive picture and structured patterns of the different methods used to model individual agent decision processes seen in a variety of applications. 

We review each paper in sequence and in some detail, providing sufficient depth in the review for a reader to understand the nature of each surveyed work. 
In particular, we focus on how data was used in the modeling process,
and in particular, in initialization, calibration and validation steps.
We attempt to draw connections among the papers using our categorization structure (i.e., by grouping them into the six categories based on the methodology used to model individual agent behavior).
Table~\ref{tb:dist_cat} shows how these survey articles are
distributed across the categories and publication years\footnote{For simplicity, we omit ``20'' and use the last two digits to denote a year. For example, ``07(2)'' stands for 2 publications in year 2007.}. 
Notice that this approach is different from the synthesis-based
approach followed by other review papers, such as, ~\cite{windrum07},
and ~\cite{macal16}, which generally draws conclusion for a
  collection of papers but does not provide sufficient detail to
  assess how data is used in these efforts. 

\begin{table}[ht]
\caption{Distribution of surveyed papers over categories and years}
\centering
\begin{tabular}{c c c}
\hline
\hline 
Category by modeling methods & Distribution in year & Total Published\\[0.5ex]
\hline
mathematical optimization based model& 01,07(2),09,10,13  & 6\\
economic model & 10, 11(2), 12, 13, 14(2), 15 & 8\\
cognitive agent model & 02, 06, 09(2), 12, 13(2), 15(2), 16(2)& 11\\
heuristic model &10, 11(2)& 3\\
statistics-based model& 07(2), 08, 09, 11, 12, 13, 14, 15, 16 & 10 \\
social influence model& 13(2), 14, 16, 17 & 5\\
\hline
Total&&43\\
\hline
\end{tabular}
\label{tb:dist_cat}
\end{table}

\subsection{Mathematical Optimization (MO) Based Models}
\label{se: mo}

The MO-based models posit that agents (e.g., farmer households) are deliberate
decision-makers who use sophisticated mathematical planning tools to assess the possible consequence of actions. While agents may encounter uncertainty, incomplete information, and constraints, their final decisions to adopt innovations are determined by concrete optimization objectives. 
The use of complex mathematical programs is commonly justified by the fact that farmer agents often consider their farming decisions in terms of economic returns. 



In a seminal paper,~\citet{berger01} developed a spatial multi-agent
mathematical programming (MP) model of diffusion of farming innovations in Chile.
Production, consumption, investment, and marketing decisions of
individual households are modeled using linear programming with the goal of maximizing expected family income subject to limited land and water assets.
Moreover, in accordance with the literature on innovation diffusion,
the model incorporates effects of past experience, as well as observed experience by peers. This is done by imposing a precondition for the MP
procedure that the net benefit is only calculated if peer adoption level
reaches the predefined threshold.
In addition to such contagion effects, agent interactions are also reflected by the feedback effects of land and water resources and return-flows of irrigation water, implemented by
coupling the economic agent decision model with hydrological components.
In simulation models, agents are cellular automata with each cell
associated with biophysical and economic attributes, such as soil
quality, water supply, land cover/land use, ownership, internal transport costs, and marginal productivity. 
\emph{These agent properties are initialized using empirical data}
derived from various data sources, including a survey that captures both agronomic and socio-economic features, and a spatial data set with information about land and water use.
\emph{Parameters were calibrated} in terms of closeness of simulation experiments and farm data at both macro and micro levels.
\emph{Validation} was then performed by regressing land use results based on the model on actual land use in the data.
Although values of the slope of this regression are reported for both
macro and micro levels, validation is incomplete. 
For instance, micro-validation is only conducted for the year when the
simulation starts due to data availability. 
Finally, the fact that validation was not conducted on data
independent from calibration is another important weakness.
Later, \citet{berger07wrm} applied his MP-based agent-based modeling approach to study the complexity of water uses in Chile. Unfortunately, that work still had the same issue on validation. 


\citet{berger07} adopted the MP-based approach to simulate soil fertility and poverty dynamics in Uganda, and analyze the impact on these of access to short-term credit and alternative technologies. 
At the heart of the model is a simulation of a farmer's decision process, crop yields, and soil fertility dynamics.
The decision model is comprised of three parts: 1) a set of possible decisions related to agriculture, such as growing crops, raising livestock, and selling and purchasing agricultural products; 2) a utility function that determines how much the decisions contribute to the farmer's objectives; and 3) links among decision variables represented by a set of equations.
Following~\citet{berger01}, a three-stage decision flow is defined that separates agent decisions into investment, production, and consumption.
Moreover, the portion of the model capturing consumption includes econometrically-specified allocation of farm and non-farm income to saving, food, and other expenditures.
Properties of the household agent, such as quantity and quality of land, labor, livestock, permanent crops, and knowledge of innovation, are sampled from empirical distributions based on limited samples.
Additional features include models of animal and tree growth, technology diffusion, demographics, and price changes.
In technology diffusion, peer influence is captured in the same manner as~\citet{berger01}, 
but notably, each agent is assigned a threshold based on household survey data.
The model was systematically validated in three steps: first, econometric models were validated for accuracy, then each component was validated independently, and finally the system as a whole.
Similar to~\citet{berger01}, validation used the same data as calibration.

\citet{schreinemachers09} studied the diffusion of greenhouse agriculture, using bell pepper in a watershed in the northern uplands of Thailand as a case study.
The work largely follows the MP-MAS (mathematical programing-based multi-agent systems) approach due to~\citep{berger01}. 
Notably, the author proposes calibrating the diffusion thresholds as described in~\citep{berger01} by using a binary adoption model (e.g., logistic regression), which is estimated from farmer survey data.
To obtain threshold values for individuals, the author first computes adoption probability for each agent based on a set of observable independent variables, and then ranks these, dividing them into the five categories of innovators due to~\citet{rogers95}.
Validation was carried by checking the value of $R^2$ associated with a regression of observed land use on its predicted value.
The proposed validation method suffers from the same limitation as other related research in using the same data for calibration and validation.


\citet{berger10} applied the MP-based approach to study the impact of several agricultural innovations on increasing profitability of litchi orchards in Northern Thailand.
Unlike~\citet{schreinemachers09} that estimated a logistic regression model to assign agents to threshold groups, they assigned thresholds randomly due to the lack of relevant data.
The model was validated using regression method as described in~\citet{schreinemachers09}, and validation suggests that the model reasonably represents aggregate agent behavior, even while individual-level behavior is not well captured.
As in prior work, calibration and validation used the same data.


\citet{alexander13} developed an agent-based model of the UK perennial energy crop market to analyze spatial and temporal dynamics of energy crop adoption. 
The model includes the interaction of supply and demand between two agent groups: farmers and biomass power plant investors. 
The farmer agents have fixed spatial locations which determine the land quality and climate that in turn impact crop yields, and decide on the selection of crops via a two-stage approach similar to~\citet{berger01}, with peer influence again modeled through a threshold function.
A farmer agent considers adoption only if the proportion of neighbors within a given radius with a positive adoption experience exceeds a threshold.
When adoption is considered, a farm scale mathematical program is used to determine the optimal selection of crops that maximizes utility as described in~\citet{alexander14}. 
Calibration of the farm scale model is either informed by empirical data or in reference to previous studies.
Validation involved checking model behaviors on simplified configurations, unit-testing of model components, and comparing simulation results against empirical data.
However, validation did not use independent data from calibration.
\subsection{Economic Models}
\label{se: ec}
Unlike the MO based models in Section~\ref{se: mo}, the \emph{economic} models use simpler rules with fewer constraints and decision variables. Particularly, agents commonly simply minimize cost, maximize profit, or, more generally, maximize personal utility. 
\subsubsection{Cost Minimization}
\citet{faber10} develop an agent-based simulation model for energy technologies, micro-CHP (combined heat and power) and incumbent condensing boilers, in competition for consumer demand.
Consumer agents are classified by housing type, which is viewed as the most important factor in determining natural gas requirements for heating units.
At each time step a consumer considers purchasing a new heating unit, and follows a three-step decision algorithm: 
1) assess if a new unit is needed, 2) scan the market for ``visible'' heating units, where ``technology awareness'' is formulated as a function of the level of advertising, market share, and bandwagon effect, and 3) each consumer chooses the cheapest technology of those that are visible.
The cost, which depends on the consumer's class, is comprised of purchase costs, subsidies, and use costs over the expected life of the technology.
Some of the parameters are calibrated using empirical data, while others are set in an ad hoc fashion.
Some validation was performed through the use of a sensitivity analysis of the variables such as market size, progress rate, and technology lifetime.
However, no explicit model validation using empirical data was undertaken.

\subsubsection{Profit Maximization}

\citet{sorda13} develop an agent-based simulation model to investigate 
electricity generation from combined heat and power (CHP) biogas plants in Germany. 
Instead of simulating farmer's individual decision whether to invest in a biogas plant, the model solves a system-wide optimization problem from the perspective of a global planner.
The model includes two types of agents: information agents, including Federal Government, Bank, Electric Utility, and Plant Manufacturer, and agents making investment decisions, including the Substrate Supplier, District, Decision-Maker, and Heat Consumer.
The core decision-making agent acts as a representative for investors in each community.
The agent chooses to invest in a biogas facility whenever sufficient resources are available and the investment yields positive net present value.
This work used multiple data sources to construct the simulation model. 
For example, plant operator guidelines and manufacturer specifications were used to obtain data about the characteristics of biogas plants.
Although the model is thus informed by real data, it is not quantitatively validated.

\subsubsection{Utility Maximization}

\citet{broekhuizen11} develop an agent-based model of movie goer behavior which incorporates social influence in movie selection decisions.
Their study investigates two types of social influence: the influence of past behavior by others, and influence stemming from preferences of an individual's friends, such as group pressure to join others in seeing a movie.
The main purpose of this work is to determine the degree to which different types of social influence impact inequality.
In their model, agent's decision-making is probabilistic and utility-driven. 
An agent first observes which movies are being shown in the marketplace with some probability.
Next, with a specified probability, an agent is selected to consider seeing a movie. 
If selected, it goes to the movie that maximizes expected utility among all those it is aware of.
Otherwise, it does not see any movie.
\emph{Utility} in this setting is a weighted sum of \emph{individual} utility, which represents the alignment between individual's preferences and movie characteristics, and \emph{social} utility which is a combination of the two types of social influence above.
Some of the model parameters are either theoretically determined or empirically calibrated, while the variability of the rest is investigated by sensitivity analysis. 
Validation involved a cross-national survey, using cross-cultural differences due to Hofstede's collectivism-individualism index to measure social influence.
While the validation is based on an independent survey study, it is largely qualitative. 


\citet{gunther11} introduce an agent-based simulation approach to support marketing activities. 
The proposed model was applied to the study of a new biomass-based fuel that would likely be introduced in Austria. 
Consumer agents are embedded in a social network, where nodes represent agents and edge weights determine the probability with which the connected agents communicate.
The authors tested several network structures, including random (Erdos-Renyi) networks, small-world networks, and so-called ``preferences-based'' networks, where connections between agents are based on geographical and cognitive proximity as well as opinion leadership. 
Each agent is characterized by preferences, geographical position, tanking behavior, how informed they are about the product, and their level of social influence. 
Agents have preferences for several product attributes: price, quality, and expected environmental friendliness, which are initialized differently based on consumer type.
Agents are geographically distributed in virtual space based on the spatial distribution of Austrian population, and their tanking behavior is a function of 
fuel tank capacity, travel behavior, and habits.
Individual information level on the innovation at hand captures the knowledge about a product, which increases as a function of interpersonal communication and exposure to marketing activities.
Influence level, on the other hand, represents an agent's expertise with the innovation and determines the amount of information received through communication. Upon interaction, an agent with lower information level learns from a more informed agent. 
Most importantly, the \emph{utility} function for agent $i$ at time $t$ is given by
$u_{i,t}=(1-Price_t)\times w_{i, 1}+Price_t\times w_{i,2}+ppq_{i,t}\times w_{i,3}+w_{i,4}$, 
where $0\leq w_{i,k} \leq 1$ and $\sum_{k=1}^4 w_{i,k}=1$, and the first and second weights pertain to price, while the last two represent how strongly agents prefer quality and how willing they are to seek renewable energy sources for fuel, respectively.
An agent is assumed to adopt if utility exceeds a specified individual threshold drawn for each agent from the uniform distribution. 
Moreover, the \emph{perceived product quality}, $ppq_i$ is assumed to gradually converge the true product quality for adopters.  
The author briefly mentions that model parameters are set in reference to a prior case study.
Apart from this, no detailed information is provided about how model parameters are actually calibrated in the setting.
Moreover, the model was only validated qualitatively with subjective expert knowledge.

\citet{holtz12} develop a utility-based agent-based model to study how farmer characteristics affect land-use changes in a region of Spain.
As relevant data are scarce, their model cannot be quantitatively calibrated and validated. 
Empirical data are used to initialize the model, deriving the initial crop distribution, and to assess the validity of the model qualitatively.
In this model, an agent's utility is formulated as a Cobb-Douglas function by multiplying four influences: gross margin, risk, labor load, and regulatory constraints.
Parameters associated with these influences differ with the types of farmers, for example, part-time, family, and business-oriented farmers would have distinct utility parameters. 
In the decision process, an agent chooses a land use pattern that maximizes its utility, where land use patterns involve a combination of crop and irrigation technology, constrained by policies.
The diffusion of irrigation technology is simulated based on the concept that the more widely used a technology is, the more likely it is to be considered by individual farmers. 
Their experiments explore the importance of each influence variable in the utility function, as well as of farmer types, by qualitatively comparing the simulation results with empirical data.

\citet{plotz14} 
propose a model for the diffusion of electric vehicles (EVs) 
to evaluate EV-related policies based driving data in Germany.
The model determines the market shares of different technologies by simulating each driving profile as both EV and conventional vehicle, choosing the option which maximizes the driver's utility, and then extrapolating these agent-level choices to aggregate market shares.
In modeling individual decisions, utility is defined as a function of \emph{total cost of ownership} (TCO), 
choice of EV brands, and individual \emph{willingness-to-pay-more} (WTPM).
The authors combined survey results with driving profiles to derive four categories of agents (adopters), and assigned each driving profiles to one of these categories.
Through simulating the \emph{plug-in hybrid electric vehicle} (PHEV) share of the market as a function of annual average \emph{vehicle kilometers traveled} (VKT) for medium-sized vehicles, the model was validated by comparing original group assignment with simulated outcomes and by examining simulated diesel market shares relative to actual values within different branches of industry.
While validation is quantitative and rigorous, it does not use independent data.
Moreover, the model does not capture social influence which is often a key aspect of innovation diffusion modeling.

\citet{mccoy14} develop an agent-based model of diffusion of electric vehicles among Irish households.
Agents representing households are located at a regular lattice space. 
They are heterogeneous as suggested by their characteristics.
Agents have two static attributes, \emph{Income Utility} (IU) and \emph{Environmental Utility} (EU), drawn independently from empirical distributions derived from a survey. 
In particular, IU is based on an agent’s social class, tenure type, and age, which are assumed to be highly correlated with income, whereas
EU is based on the agent's past adoption of energy efficiency technologies and their attitude toward the environment. 
Each agent $i$ has a unique threshold, $\theta_i$, drawn from a distribution that is negatively correlated to IU, and adopts if 
$U_i(t)\geq \theta_i \text{ and } t\times crit \geq rand(0, 1)$, 
where, $crit$ is decimal value that is used to account for inertia that exists in early stage of technology adoption, 
while utility $U_i(t)$ is defined as
$U_i(t)=\alpha_i IU_i+\beta_iEU_i+\gamma_iG_i(t)+\delta_iS(t)$,
where, $IU$ represents individual's preferences, $G$ is social influence, and $S$ is social norms, and $\alpha_i+\beta_i+\gamma_i+\delta_i=1$. 
To allow these parameters to vary by agent, the authors specify four distinct consumer groups with different preferential weighting schemes. 
Although the agents in the simulation are initialized using empirical distributions, key parameters in the decision model are not derived empirically but are based on the authors' assumptions. 
Additionally, no rigorous validation is provided.

\citet{palmer15} developed an agent-based model of diffusion of solar photovoltaic (PV) systems in the residential sector in Italy.
The \emph{utility} of agent $j$ is defined as the sum of four weighted partial utilities, i.e., $U(j)=w_{pp}(sm_j)\cdot u_{pp}(j)+w_{env}(sm_j)\cdot u_{env}(j)+w_{inc}(sm_j)\cdot u_{inc}(j)+w_{com}(sm_j)\cdot u_{com}(j)$, where $\sum_k{w_k(sm_j)}\\
=1$ for $k\in K: \{pp, env, inc, com\}$ and $w_k(sm_j), U(j)\in [0, 1]$.
From left to right the partial utilities are: (1) payback period of the investment, (2) environmental benefits, (3) household income, and (4) social influence.
An agent chooses to invest in PV if its total utility exceeds an exogenously specified threshold.
Thresholds above vary by agent's demographic and behavioral characteristics, $sm_j$.
The four partial utilities are derived from empirical data. 
Specifically, the payback period is estimated based on investment costs, local irradiation levels, government subsidies, net earnings from generating electricity from the system vs.\ buying it from the grid, administrative fees, and maintenance costs.
The environmental benefit is based on an estimate of reduced $CO_2$ emissions saved.
Household income is estimated based on household demographics, such as age, level of education, and household type.
Finally, social influence is captured by the number of neighbors of a household within its social network who have previously adopted PV.
The social network among agents is generated according to the small-world model~\citep{watts98},
modified to account for socio-economic factors. 
The model parameters are calibrated by trying to match simulated adoption with the actual aggregate residential PV adoption in Italy over the 2006-2011 period.
The model is then applied to study solar PV diffusion in Italy over the 2012-2026 period.
However, no quantitative validation is offered.

\subsection{Cognitive Agent Models}
\label{se:ca}
While both MO-based (Section~\ref{se: mo}) and economic (Section~\ref{se: ec}) models elaborate economic aspects of the decision process and integrate simple threshold effects, cognitive agent models aim to explicitly model how individuals affect one another in cognitive and psychological terms, such as opinion, attitude, subjective norm, and emotion. This category includes the \emph{Relative Agreement Model}, the \emph{Theory of Planned Behavior}, the \emph{Theory of Emotional Coherence}, and the \emph{Consumat Model}.

\subsubsection{Relative Agreement Model}

The \emph{Relative Agreement Model} belongs to a class of opinion dynamics models~\citep{hegselmann02} and addresses how opinion and uncertainty are affected by interpersonal interactions. Seminal work is due to~\citet{deffuant00}, who investigate how the magnitude of thresholds, with respect to attitude difference, leads to group opinion convergence and extremeness. The relative agreement model is often known as ``\emph{Deffuant model}'' in the literature. 




\citet{deffuant02} design an agent-based model to simulate organic farming conversion in France. 
To model impact of interactions on the individual decision, they relied on the Deffuant model in which both opinion and uncertainty are continuous variables.
In the diffusion model, farmer agent has an ``interest'' state with three possible values: \emph{not-interested}, \emph{uncertain}, and \emph{interested}. 
The actual value is based on the agent's opinion (represented as a mean value and confidence) and economic consideration.
The value of the interest state depends on the position of the global opinion segment compared to a threshold value.
Agent changes opinion after discussing with peers using a variant of the \emph{Relative Agreement} algorithm~\citep{deffuant02a}. %
The farmers send messages containing their opinions and information, following a two-stage diffusion model of~\citet{vale95}, mediated by a network generated according to the~\citet{watts98} model.
These impact opinions of the recipients as a function of opinion similarity, as well as confidence of the sender, with more confident opinions having greater influence.
In addition, if the farmer agent 
is ``interested'' or ``uncertain'', he performs an evaluation of the economic criterion, and if he remains interested, he requests a visit from a technician.
After this visit, the economic criterion is evaluated again under reduced uncertainty.
Finally, the adoption decision is made when the farmer has been visited by a technician and remains ``interested'' for a given duration.

Many model parameters governing the decision and communication process are not informed by empirical data.
The authors tested the sensitivity of the model by varying these variables, including the main parameters of the dynamics, the parameters of the initial opinion distribution average number of neighborhood and professional links, and variations of the institutional scenario. 
Within this parametric space, they aimed to identify parameter zones that are compatible with empirical data. 
For each parameter configuration, the authors defined two error measures: the adoption error
and the error of proximity of adopters to the initial organic farmers. 
A decision tree algorithm was then used to find the parameter zones where the simulated diffusion has an acceptable performance. 
While this sensitivity analysis step can be viewed as model calibration, it is distinct from classical calibration which aims at finding a single best parameter configuration.
The model was not validated using independent data.

\leaveout{
The ``Deffuant model'' is just sub model of the entire simulation system. Its role is to define how agent interactions modify individual opinions. On the other hand, the decision-making process relies on an analog of typical farmer decision-making process as summarized from the field study represented by a sequence of predefined actions.
The method is intuitive, but one would need to define a large set of parameters and make a series of probabilistic assumptions on individual decision-making and communication rules. 
Estimating these parameters seems challenging but also unnecessary, since a lot of detailed information about how farmers behave and communicate would be required. 
Probably due to this inefficiency, the ``Deffuant model'' is often coupled with other decision models, such as the Theory of Planned Behavior (TPB)~\citep{ajzen91}, which provides with us a alternative and simpler way to organize agent decision-making process in form of structural equations. Another big advantage of using TPB based models is that one can estimate the parameters through regression analysis directly from well-established survey methods. 
}

\subsubsection{Theory of Planned Behavior}
The \emph{Theory of Planned Behavior (TPB)} postulates that an individual's \emph{intention} about a behavior is an important predictor of whether they will engage in this behavior~\citep{ajzen91}.
As a result, the theory identifies three attributes that jointly determine intention: \emph{attitudes}, \emph{subjective norms}, and \emph{perceived behavioral control}.
The relative contribution for each predictor is represented by a weight which is often derived empirically using regression analysis based on survey data. 

\leaveout{
A number of closely related models are common in the study of information technology adoption.
For example, 
the Theory of Reasoned Action (TRA) has been used in empirical studies to predict user acceptance of information technology, in which intention is modeled as a weighted sum of attribute and subjective norm. 
\citet{muduganti05} design a basic computational TRA model of
user acceptance of information technology, and conducted experiments by varying the parameters such as the relative importance of the attitude and subjective norm components, population size, and threshold values. 
TPB can be viewed as a refinement of TRA by incorporating the perceived behavioral control, which is an important consideration in diffusion modeling.
}

\citet{kaufmann09} build an agent-based simulation model on TPB to study the diffusion of organic farming practices in two New European Union Member States.  
Following the TPB methodology, each agent is characterized by three attributes: the attitude $a_i$, subjective norm $s_i$, and perceived behavioral control $p_i$, each ranging from -1 (extremely negative) to +1 (extremely positive). 
The intention $I_i$ is defined as 
$I_i=w_i^aa_i+w_i^ss_i+w_i^pp_i$, 
where $w_i^a, w_i^s, w_i^p$ are relative contribution toward intention.
The weights for non-adopters and adopters are derived separately using linear regressions based on the survey data.
If an agent's intention exceeds a threshold $t$ it adopts, and does not adopt otherwise.
The threshold is obtained from survey data as the average intention of non-adopters who have expressed a desire to adopt.
In the simulation model, social influence is transmitted among network neighbors in each time step in a random order.
Specifically, when one node speaks to another, the receiver shifts its subjective norm closer to the sender's intention, following the relative agreement framework.
Social networks are generated to reflect small-world properties~\citep{watts98} and a left-skewed degree distribution~\cite{noble04}, with specifics determined by a set of parameters, which are set based on survey data (such as the average degree). 
While empirical data is thus used to calibrate parameters of the model, no quantitative validation was provided.

\citet{schwarz09} propose an agent-based model of diffusion of water-saving innovations, and applied the model to a geographic area in Germany. 
Agents are households with certain lifestyles, represented by demographic and behavioral characteristics. 
They use two different decision rules to determine adoption: a cognitively demanding decision rule representing a \emph{deliberate} decision and a simple decision \emph{heuristic}.
The particular decision rule to use is selected based on the agent's type and technology category.
The deliberate decision-making algorithm is based on multi-attribute subjective utility maximization that integrates attitude, social norm, and perceived behavioral control.
The heuristic decision rule makes decisions in greedy order of evaluation criteria based on innovation characteristics and social norms.
Finally, if no clear decision can be made, agents imitate their peers, who are defined through a variation of a small-world network~\citep{watts98} which captures spatial proximity and lifestyle affinity in determining links among agents.
The model was calibrated using data from a survey according to the framework of the Theory of Planned Behavior~\citep{ajzen91}, with the importance of different decision factors derived by structural equation models or linear regressions for lifestyle groups. 
The model was validated using \emph{independent} market research data at the household level. 
In addition, due to the lack of independent aggregated diffusion data, results of the empirical survey were used for validation.
\citet{sopha13} present an agent-based model for simulating heating system adoption in Norway. 
Their model extends TPB to consider several contributing factors, such as household groups, intention, attitudes, perceived behavioral control, norms, and perceived heating system attributes. 
Households are grouped using cluster analysis based on income level and basic values available in the survey data to approximate the influence of lifestyle on attitudes towards a technology. 
Attribute parameters are then estimated using regressions for each household cluster based on the household survey. 
Moreover, motived by the meta-theory of consumer behavior~\citep{jager00a}, the model assumes that a household agent randomly follows one of four decision strategies: repetition, deliberation, imitation, and social comparison, in accordance with empirical distribution based on survey data.  
Notably, this model is validated using \emph{independent} data that is not used for calibration, examining how well simulation reproduces actual system behavior at both macro and micro level.


\citet{rai15} develop an empirically grounded agent-based model of residential solar photovoltaic (PV) diffusion to study the design of PV rebate programs.
The model is motived by TPB and assumes that two key elements determine adoption decision: attitude and (perceived) control.
The authors calibrate population-wide agent attitudes using survey data and spatial regression.
Following the opinion dynamics model in~\citet{deffuant02a}, at each time-step, agents' attitudes about the technology and their uncertainties are adjusted through interactions with their social network neighbors following the relative agreement protocol.
Social influence is captured by households situated in small-world networks, with most connections governed by geographic and demographic proximity.
In the ``control'' module, an agent $i$ compares its perceived behavioral control $pbc_i$ with the observed payback at the current time period $PP_{it}$. 
Then, if the agent exceeds its attitude threshold, it adopts when $PP_{it}<pbc_i$.
$pbc_i$ for each agent $i$, is calculated as a linear sum of financial resources, the amount of sunlight received, and the amount of roof that is shaded, while
$PP_{it}$ is calculated based on electricity expenses offset through the use of the solar system, the price of the system, utility rebates, federal investment tax credit, and annual system electricity generation.
The six model parameters used to specify the social network, opinion convergence, the distribution of the behavioral control variable, and the global attitude threshold value were calibrated by an iterative fitting procedure using historical adoption data. 
The model was first validated in terms of predictive accuracy, comparing predicted adoption with empirical adoption level for the time period starting after the last date for the calibration dataset.
Moreover, temporal, spatial, and demographic validation were conducted. 
However, validation was focused on aggregate (macro), rather than individual (micro) behavior.

\citet{jensen16}
develop an agent-based model to assess energy-efficiency impacts of an air-quality feedback device in a German city.
A household agent makes two decisions: whether to adopt a feedback device and whether to practice a specific energy-saving behavior.
The model involves simulating both the adoption of the feedback device and the heating behavior respectively.
Two diffusion processes are connected based on the observation that the feedback device changes an agent's heating behavior, and eventually will form a habit.
In the simulations, household agents are generated based on marketing data on lifestyle, and initial adopters of the heating behavior are selected based on a survey.
The adoption of an energy-efficient heating behavior is triggered by external events, whose rate is estimated by historical data using Google search queries.
Their survey reveals that both information and social influence drive behavior adoption.
This insight is integrated into a decision-making model following the theory of planned behavior (TPB), in which information impacts the agent's attitude in each simulation step.
On the other hand, the diffusion model of the feedback device is an adaptation of an earlier model also based on TPB.
An adopter of the device is assumed to adopt the desired heating behavior with a fixed probability, which is informed by an empirical study. 
The space of model parameters is reduced by applying a strategy called ``pattern-oriented modeling'', which refines the model by matching simulation runs with multiple patterns observed from empirical data
\citep{grimm05}.
In their experiments, the authors calibrated several different models using empirical data and aimed to quantify the effect of feedback devices by comparing results generated by these models. 
However, no rigorous model validation is presented.

\subsubsection{Theory of Emotional Coherence}
When it comes to explaining and predicting human decisions in a social context, some computational psychology models also take emotional factors into account, which are often neglected by TPB-based models.   
\citet{wolf12} propose an agent-based model of adoption of electric vehicles by consumers in Berlin, Germany, based on the \emph{Theory of Emotional Coherence (TEC)}. 
The parameters of the model were derived based on empirical data from focus groups and a representative survey of Berlin's population. 
In particular, the focus group provided a detailed picture of people's needs and goals regarding transportation; the survey was designed to generate quantitative estimates of the beliefs and emotions people associate with specific means of transportation.
The attributes of the agents include age, gender, income, education, residential location, lifestyle, and a so-called social radius, and are obtained based on the survey data.
The social network structure is generated by similarities between these characteristics following the theory of \emph{homophily}~\cite{mcpherson01}; specifically, the likelihood of two individuals communicating with one other is a function of their similarity in terms of demographic factors.  
To validate the predictions made by the model, the authors regressed empirical data related to actual transportation-related decisions (e.g., weekly car usage) from the survey on the activation parameters resulting from simulations. 
However, validation did not use independent data.

\subsubsection{Consumat Model}
The Consumat Model is a social psychological framework, in which consumer agents switch among several cognitive strategies---commonly, \emph{comparison}, \emph{repetition}, \emph{imitation}, and \emph{deliberation}---as determined by need satisfaction and their degree of uncertainty~\citep{jager00b}. 
\citet{schwoon06} uses an agent-based model (ABM) to simulate possible diffusion paths of fuel cell vehicles (FCVs), capturing complex dynamics among consumers, car producers, and filling station owners. 
In their model, the producers offer heterogeneous but similar cars, deciding in each period whether to change production to FCVs.
Consumers have varying preferences for car attributes, refueling needs, and social influence factors.
Although in a typical consumat approach~\citep{janssen02}, consumers follow one of four cognitive strategies 
on the basis of their level of need satisfaction and uncertainty, the author rules out repetition and imitation and argues that need satisfaction is rather low in their case.
The consumer agent is assumed to maximize total expected utility, which is expressed as a function of car price, tax, the closeness between preferences and car characteristics, social need, as determined by the fraction of neighbors adopting each product type, and availability of hydrogen.  
In the model, individual preferences may evolve with time to be more congruent with the ``average car'', as determined by a weighted average of attributes of cars sold in the previous period, where weights correspond to market shares.
The model is calibrated by trying to match main features of the German auto market.
The network structure governing social influence is assumed to form a torus.
The model does not attempt quantitative validation.

\subsubsection{The LARA Model}
LARA is the short for \emph{Lightweight Architecture for boundedly Rational Agents}, a simplified cognitive agent architecture designed for large-scale policy simulations~\citep{briegel12}.
Comparing with existing complex psychological agent frameworks, LARA is more generalizable and easier to implement. 
We review two recent efforts motivated by the LARA architecture and grounded in empirical data.

\citet{krebs13} develop an agent-based model to simulate individual's provision of neighborhood support in climate change adaptation.
In their model, agents are assigned to lifestyle groups and initialized using spatial and societal data.
Motivated by LARA, an agent makes a decision in one of three modes: deliberation, habits, and exploration.
In deliberation, an agent compares and ranks available options in terms of utility, which is the weighted sum of four goals: striving for effective neighborhood support, being egoistic, being altruistic, and achieving social conformity.  
The goal weights, which are different among lifestyle groups, are set based on expert ratings and the authors' prior work. 
A probability choice model is used to choose the final option when multiple better options are available. 
An agent acts in deliberation mode if no experience is available (habitual behavior is not possible) and shifts to an exploratory mode with a predefined small probability. 
The network in which the agents are embedded is generated using lifestyle information.
Simulation runs for an initial period from 2001 to 2010 provide plausible results on behavioral patterns in cases of weather changes. 
From 2011 to 2020, the authors examine the effects of two intervention strategies that mobilize individuals to provide neighborhood support.
Some model parameters remain uncalibrated, and the entire model is not validated due to a lack of empirical data at the macro level.  

\citet{krebs15}
develop an agent-based spatial simulation of adoption of green electricity in Germany.
Each agent represents a household deciding to select between ``green'' and ``gray'' energy providers.
Every agent is characterized by its geographical location and lifestyle group. 
Agents are initialized and parameterized by empirical data from surveys, psychological experiments, and other publicly available data. 
Following LARA, agents are assumed to make decisions either in a deliberative or habitual mode.
Default agent behavior is habitual, and the agent transitions to a deliberative mode when triggered by internal and external events, such as a price change, personal communication, cognitive dissonance, need for cognition, and media events.
An agent chooses an action that maximizes utility, which is a weighted sum of four goals: ecological orientation, economic orientation, social conformity, and reliability of provision.
The goal weights depend on the lifestyle group and are derived from a survey and expert rating~\citep{ernst16}.
An artificial network that connects the agents is generated based on lifestyle and physical distance~\citep{ernst16}. 
Once an agent decides to adopt green electricity, it chooses a service brand that is already known.
The diffusion of the awareness of the brand is characterized by a simple word-of-mouth process.
Validation focuses on two state variables of agent behavior: selected electricity provider and awareness of the brand, which involves comparing simulation results with historical data both temporally and spatially starting from aggregate to the individual level.
Unfortunately, validation was not conducted using independent data.

\subsection{Heuristic Models}

\label{se:hr}
Heuristic adoption models are often used when modelers are not aware of any established theories for agent decision-making in the studied application. 
These models tend to give us an impression of being ``ad-hoc'', since they are not built on any grounded theories. 
More importantly, unlike the cognitive agent models such as the theory of planned behavior, there is no established or principled means to estimate model parameters. 
Therefore, model parameters are often selected in order to match simulated output against a realistic adoption level. Although heuristic-based model appears to be an inaccurate representation of agent decision-making, they are easy to implement and interpret.


\citet{vliet10} make use of a \emph{take-the-best} heuristic to model a fuel transportation system 
to investigate behavior of fuel producers and motorists in the context of diffusion of alternative fuels. 
In the model, producers' plant investment decision is determined by simple rules, and the same plant can produce multiple fuel types.
Motorists are divided into several subgroups, each having distinct preferences.
Each motorist is assumed to choose a single fuel type in a given year.
Each fuel is assigned four attributes: driving cost, environment, performance, and reputation.
Motorist preferences in the model are represented by two factors: 1) \emph{priorities}, or the order of perceived importance of fuel attributes, and 2)  \emph{tolerance} levels, which determine how much worse a particular attribute of the corresponding fuel can be compared to the best available alternative to maintain this fuel type under consideration.
The decision heuristic then successively removes the worst fuel one at a time in the order of attribute priorities.
Due to the difficulty of obtaining actual preferences of motorists, the authors used the Dutch consumer value dispositions from another published model in literature as a proxy to parameterize the model.
However, the model was not rigorously calibrated or validated using empirical data.

\citet{zhao11} propose a two-level agent-based simulation modeling framework to analyze the effectiveness of policies such as subsidies and regulation in promoting solar photovoltaic (PV) adoption.
The lower-level model calculates payback period based on PV system electricity generation and household consumption, subsidies, PV module price,
and electricity price. 
The higher-level model determines adoption choices as determined by attributes which include payback period, household income, social influence, and advertising.
A pivotal aspect of the model is the \emph{desire} for the technology (PV), which is formulated as a linear function of these four factors, and an agent adopts if the desire exceeds a specified threshold.
Survey results from a prior study were used to derive a distribution for each factor, as well as the membership function in a fuzzy set formulation.
The agents in the model were initialized using demographic data, along with realistic population growth dynamics based on census data.
Moreover, calibration of threshold value was conducted to match simulated annual rate of PV adoption with historical data.
However, the model was not quantitatively validated using independent data.

A more complex TOPSIS (\emph{Technique for Order Preference by Similarity to Ideal Solution}) model is a decision heuristic which selects an option from several alternatives that is the closest to the ideal option and the farthest from the worst possible option. 
\citet{kim11} present agent-based automobile diffusion model using a TOPSIS approach to simulate market dynamics upon introduction of a new car in the market.
%
The model integrates three determinants of purchasing behavior: (1) information offered by mass media, (2) relative importance of attributes to consumers, and (3) social influence. 
Individual agents rank products by considering multiple product attributes and choosing a product closest to an ideal.
A survey was conducted to estimate consumers' weights on the car attributes and the impact of social influence.
In the simulations, diffusion begins with innovators who try out new products before others; once they adopt, their social network neighbors become aware of these decisions, with some deciding to adopt, and so on.
A \emph{small-world} network structure was assumed for this virtual market, and choices of rewiring and connectivity were determined by the model calibration step through comparing simulated results with historical monthly sales volumes of three car models.
However, the model was not validated using independent data.

\subsection{Statistics-Based Models}
\label{se: st}
Statistics-based models rely on statistical methods to infer relative contribution of observable features towards one's decision whether to adopt. The estimated model is then integrated into an ABM. We review three subcategories of statistics-based methods for agent-based models of innovation diffusion: \emph{conjoint analysis}, \emph{discrete choice models}, and \emph{machine learning}. 

\subsubsection{Conjoint Analysis}
Conjoint analysis is a statistical technique used in market research to determine how much each attribute of a product contributes to consumer's overall preference. This contribution is called the \emph{partworth} of the attribute. Combining with feature values of innovation obtained from field study, one can construct a utility function accordingly. 

\citet{garcia07} utilize conjoint analysis to instantiate and calibrate an agent-based marketing model using a case study of diffusion of Stelvin wine bottle screw caps in New Zealand. 
With a particular emphasis on validation, the overall work follows \citet{carley96}'s four validation steps: \emph{grounding}, \emph{calibration}, \emph{verification}, and \emph{harmonizing} (the latter not performed, but listed as future work) to properly evaluate the model at both micro and macro levels. 
The model includes two agent types: wineries and consumers. 
In each period the wineries set the price, production level, and attributes of screw caps as a function of consumer demand.
Consumers, in turn, make purchase decisions following their preferences.
The model is calibrated using conjoint analysis, inferring \emph{partworths} which determine consumer preferences in the model.
Aggregate stylized facts were then replicated in the verification step.
The work emphasizes the value of calibration, but pays less attention to validation, which is merely performed at a face level rather than quantitatively. 

\citet{vag07} presents a \emph{dynamic} conjoint method that enables forecasts of future product preferences.
The consumer behavior model considers many factors, including social influence, communication, and economic motivations.
The author surveys behavior of individuals, such as their communication habits, and uses conjoint analysis to initialize preferences in the ABM.
Notably, in this model agent priorities depend on one another, and the resulting social influence interactions may lead to large-scale aggregate shifts in individual priorities.
To demonstrate the usability of their model, the study utilized empirical data on product preferences (in this case, preferences for mobile phones), consumer habits, and communication characteristics in a city in Hungary.
Calibration of this model was only based on expert opinion and comparative analysis, rather than quantitative comparison with real data, and no quantitative validation was performed.

\citet{zhang11} develop an agent-based model to study the diffusion of eco-innovations, which in their context are alternative fuel vehicles (AFVs).
The model considers interdependence among the manufacturers, consumers, and governmental agencies in the automotive industry.
The agents representing manufacturers choose engine type, fuel economy, vehicle type, and price, following a simulated annealing algorithm, to maximize profit in a competitive environment until a Nash equilibrium is reached~\citep{michalek04}.  
The consumer agents choose which products to purchase.
The partworth information in the utility function was derived by \emph{choice-based conjoint} analysis using an empirical survey from~\citet{garcia07}. 
In particular, the probability of a consumer choosing a vehicle is formulated as a logit function of vehicle attributes, word-of-mouth, and domain-specific knowledge. 
The utility is modeled as a weighted sum of attributes, and parameters/partworth are estimated using hierarchical Bayes methods. 
The agent acting as ``government'' chooses policies aimed at influencing the behavior of both manufacturers and consumers.
Model calibration involved conjoint analysis.
However, the authors found that the ABM tended to overestimate the market shares of alternative fuel vehicles, which motivated them to adjust model parameters and to linearize the price parthworth in order to ensure that aggregate demand decreases with the price.
Like~\citet{garcia07}, the authors follow the four steps of validation~\citep{carley96}.
However, validation does not use data independent from calibration.

\citet{lee14} introduce an agent-based model of energy consumption by individual homeowners to analyze energy policies in the U.K. 
The model utilizes historical survey data and choice-based conjoint analysis to estimate the weight of a hypothetical utility function, defined as the weighted sum of attributes. 
In the simulation, moving and boiler break-down events are assumed to trigger a decision by the household agent.
In this case, a particular alternative is selected if its utility is higher than all other alternatives as well as the status quo option.
The model was populated with initial data based on a survey conducted in the U.K., and each agent was matched to a household type which can be further mapped to energy demand using energy consumption estimates.
The authors then combined energy demand with fuel carbon intensity to determine annual household emissions.
The model was calibrated by adjusting the weights in the decision model to match historic installation rates from 1996 to 2008 for loft insulation and cavity wall insulation.
The model was not validated using independent data.

\citet{stummer15} devise an agent-based model to study the diffusion of multiple products.
Each product is characterized by a number of attributes determined by expert focus group discussion. True performance of each product attribute is unknown to consumers, and each agent, therefore, keeps track of the distribution of attribute values based on information previously received.
This information is updated based on interactions with peers, advertising, or direct experience. 
Consumer agent behavior is governed by a set of parameters that capture heterogeneous preferences and mobility behavior. 
Agents have additive multi-attribute utilities, the weights of which were obtained from survey data using conjoint analysis.
The authors adapt the preferential attachment algorithm introduced by \citep{barabasi99b} to generate networks in which the attachment probability depends on both node degree and geographic distance between nodes. 
Network parameters were determined by taking into account additional information revealed in the consumer survey, such as the number of social contacts and communication frequency.
An agent decides to purchase a product which maximizes utility.
The model defines each advertising event to communicate a set of product attributes, which either increase product awareness or impact customer preferences.
The model was validated extensively following~\citep{knepell93}, including conceptual validity, internal validity, micro-level external validity, macro-level external validity, and cross-model validity.
The weakness of validation, however, is that it is only performed as an in-sample exercise without using independent data.

\subsubsection{Discrete Choice Models}
The \emph{discrete choice} modeling framework, which originates in econometrics, is used to describe, explain, and predict agent choices between two or more discrete alternatives~\citep{train09}. The approach has a wide range of applications, and we review several efforts targeted specifically at innovation diffusion.

\citet{galan09} design an agent-based model to analyze water demand in a metropolitan area. 
This model is an integration of several sub-models, including models of urban dynamics, water consumption, and technological and opinion diffusion.
The opinion diffusion model 
assumes that an agent's attitude towards the environment determines its water consumption, i.e., an non-environmentalist would use more water than an environmentalist. 
Accordingly, it is assumed that each agent can be in two states: \emph{environmentalist} (E) or \emph{non-environmentalist} (NE). 
The choice of a state depends on the agent's current state, the relative proportion of E and NE neighbors, and an exogenous term measuring the pressure towards E behavior.
Transition probabilities between states E and NE are given in form of \emph{logistic} functions.
However, rather than using empirical data to estimate parameters of these functions, the authors parameterized the behavior diffusion model with reference to models in prior literature for other European cities.
To determine adoption of water-saving technology, the opinion diffusion model is coupled with the technological diffusion model, which is implemented by a simple agent-based adaptation of the Bass model following~\citep{borshchev04}.
The model was validated qualitatively by domain experts, quantitatively calibrated based on the first quarter of 2006, and validated by comparing the model with actual adoption in the following two quarters.
The authors demonstrate that simulation results successfully replicate the consequence of a water-saving campaign on domestic water consumption. 

\citet{dugundji08} propose a computational model that combines econometric estimation with agent-based modeling to study the adoption of transportation options for households in a city in Netherlands. 
The presented discrete choice modeling framework aims to address interactions within different social and spatial network structures.
Specifically, agent decision is captured using a \emph{nested logit} model, which enables one to capture observed and unobserved behavior heterogeneity.
Feedback effects among agents are introduced by adding a linear term (a so-called \emph{field variable}) that captures proportions of an agent's neighbors making each decision to each agent's utility function. 
Because survey data on interactions between identifiable individuals was unavailable, this term only captured aggregate interactions among socioeconomic peers.
The authors investigated simulated transition dynamics for the full model
with two reference models: the first a nested logit model with a global field variable only and a fully connected network, and the second a multinomial logit model which is a special case to the full model. 
They found that simulated dynamics differ dramatically between the models.
Given this lack of modeling robustness, no further validation was undertaken.  

\citet{tran12} develops an agent-based model to investigate energy innovation diffusion. 
Agent behavior in this model is determined by the relative importance of technology attributes to the agents, and social influence.
Social influence, in turn, takes two forms: indirect influence coming from the general population, and direct influence of social network neighbors.
The author drew on ABM studies in the marketing literature, and formulated the adoption model as
$Prob(t)=1-(1-P_{ij})(1-Q_{ij})^{K_{ij}}$,
where $P_{ij}$ captures individual choice using a discrete choice model of consumer decision-making, in which an agent's utility is defined as an inner product of coefficients and attributes. Coefficients are a random vector, with distribution different for different agents, capturing preference heterogeneity.
$Q_{ij}$ and $K_{ij}$ is the indirect and direct network influence, respectively, captured as a function of the number of adopters at decision time. 
While the model was evaluated using simulation experiments, and the nature of the model makes it well suited for empirically grounded parameter calibration, it was not in actuality quantitatively calibrated or validated using empirical data.

\subsubsection{Machine Learning Models}
\emph{Machine learning} (ML) is a sub-area of computer science that aims to develop algorithms that uncover relationships in data.
Within a supervised learning paradigm which is of greatest relevance here, the goal is further to develop models that accurately predict the value of an outcome variable for unseen instances.
To do so, a computer program is expected to recognize patterns from a large set of observations, referred to as a \emph{training} process that is grounded in statistical principles and governed by intelligent algorithms, and make predictions on new, unseen, instances.
This category of methods has recently drawn much attention in academia and industry due to tremendous advances in predictive efficacy on important problems, such as image processing and autonomous driving.
Combining machine learning with agent-based modeling seems promising in the study of innovation diffusion since the two can complement each other. 
The former is specialized in building a high-fidelity predictive models, while the latter captures dynamics and complex interdependencies.
Of particular relevance to combining ML and ABM is the application of machine learning to model and predict human behavior.
Interestingly, relatively few attempts have been made to date to incorporate ML-based models of human behavior within ABM simulations.

\citet{sun13} develop an agent-based model that features Bayesian belief networks (BBNs) and opinion dynamics models (ODMs) to model land-use dynamics as they relate to payments for ecosystem services (PES).
The decision model of each household is represented using a BBN, which were calibrated using survey data and based on discussions with relevant stakeholders, and incorporate factors such as income and land quality.
Social interactions in decision-making are captured by ODM. 
The modeling framework was applied to evaluate China's Sloping Land Conversion Program (SLCP), considered among the largest PES programs.
SLCP was designed to incentivize reforestation of land through monetary compensation.
In their model, farmers make land-use decisions whether or not to participate in the SLCP program based on \emph{internal} beliefs and \emph{external} influences.
External influences adjust internal beliefs cumulatively using a modified Deffuant model~\citep{deffuant02a} within a community-based small-world social network. 
Initial model structures were obtained using a structural learning algorithm, with results augmented using qualitative expert knowledge, resulting in 
a pseudo tree-augmented naive Bayesian (TAN) network. 
The final BBN model was validated by using a sensitivity analysis, and measuring prediction accuracy and area under the curve (AUC) of the receiver operating characteristics (ROC) curve on a holdout test data set at both household and plot level. 
A crucial limitation of this work is that only the BBN model was carefully validated; the authors did not validate the full simulation model at either the micro or macro levels.

\citet{zhang16} propose a data-driven agent-based modeling (DDABM) framework for modeling residential rooftop solar photovoltaic (PV) adoption in San Diego county. 
In this framework, the first step is to use machine learning to calibrate individual agent behavior based on data comprised of individual household characteristics and PV purchase decisions.
These individual behavior models were validated using cross-validation methods to ensure predictive efficacy on data not used for model calibration, and were then used to construct an agent-based simulation with the learned model embedded in artificial agents.
In order to ensure validation on independent data, the entire time series data of individual adoptions was initially split along a time dimension.
Training and cross-validation for developing the individual-level models were performed only on the first (early) portion of the dataset, and the aggregate model was validated by comparing its performance with actual adoptions on the second, independent time series, into the future relative to the calibration data set.
The authors thereby rigorously demonstrate that the resulting agent-based model is effective in forecasting solar adoption both at the micro and macro levels.
To our best knowledge, this work proposed the first generic principled framework that combines ML and ABM in study of innovation diffusion. 
Unlike most ABM studies we have reviewed, DDABM has the following features:
1) it does not make any assumptions on the structural features of social network, relying entirely on a data-driven process to integrate most predictive spatial and social influence features into the individual adoption model;
2) it does not rely on matching simulated dynamics with the empirical observations to calibrate the model, but instead parameterizes the model through a far more efficient statistical learning method at the level of individual agent behavior; and
3) validation is performed on \emph{independent} data to evaluate the predictive effectiveness of the model. Moreover, validation is not only done at the macro-level by comparison with actual adoption traces, but also at the micro-level by means of the simulated \emph{likelihood ratio} relative to a baseline model. 
To further justify the usefulness of ML-base approach, \citet{zhang16} actually implement and compare their model with another agent-based model of rooftop solar adoption developed by~\citep{palmer15}, with parameters calibrated on the same dataset following the general aggregate-level calibration approach used by them.
The result is very revealing, as it strongly suggests that aggregate-level calibration is prone to overfit the model to data, an issue largely avoided by calibrating individual agent behavior.

\subsection{Social Influence Models}
\label{se:si}
Our last methodological category covers several models looking specifically at social influence. These models are quite simple, abstract, but prevalent in the theoretical study of innovation diffusion.
Our purpose of discussing these is that there have been several recent efforts to calibrate these models using empirical data.

After analyzing an adoption dataset of Skype, \citet{karsai14} develop an agent-based model to predict diffusion of new online technologies.
Specifically, agents in their model are characterized by three states: \emph{susceptible} (S), \emph{adopter} (A), and \emph{removed} (R).
Susceptible refers to people who may adopt the product later.
Adopter agents have already adopted.
Finally, removed are those who will not consider adopting the product in the future again.
The transition from S to A is regulated by \emph{spontaneous adoption} and \emph{peer-pressure}, from A to S by \emph{temporary termination}, and from A to R by \emph{permanent termination}, each of which is parametrized by a constant probability which is identical for all users.
While some parameters, such as average degree and temporary termination probability, are estimated directly from observations, the remaining parameters are determined by simultaneously fitting the empirical rates using a bounded nonlinear least-squares method.
The model is fit over a 5-year training period, and validation uses predictions over the last six months of the observation period. 
However, validation is somewhat informal, since the predictability of the model is evaluated on a part of the training data and there is no validation of micro-behavior. 
In a later work using the same Skype data,~\citet{karsai16} 
develop a threshold-driven social contagion model with only two states: susceptible and adopted. 
In addition, the model assumes that some fraction of nodes never adopt.
The authors calibrated the value of this fraction by matching the size of the largest component of adopters given by the simulations with real data.
In addition, the model assumes that susceptible nodes adopt with a constant probability, which is informed by empirical analysis.
In their simulations, nodes have heterogeneous degrees and thresholds, which follow empirical distributions.
However, validation was not performed using independent data. 

\citet{herrmann13} present two agent-based models of diffusion dynamics in online social networks.
The first ABM is motivated by the Bass model, but time is discretized and each agent has two states: \emph{unaware} and \emph{aware}. At each time step, an unaware agent changes state to aware as a function of two triggers:
innovation arising from exogenous sources, such as advertising, and imitation, which comes from observing decisions by neighbors. 
The second model termed the \emph{independent cascade} model, originating from~\citet{goldenberg01}, has a similar structure to the agent-based Bass model, except that the imitation effect is formulated as a single probability with which each aware neighbor can independently change the state of an agent to \emph{aware}.
The author applied the two models in parallel to four diffusion data sets from Twitter, and calibrated parameters using actual aggregate adoption paths.  
Notably, validation is only performed at macro-level as an in-sample exercise, and shows that the two models behave similarly.

Using historical diffusion data of Facebook apps, \citet{trusov13} introduce an approach that applies Bayesian inference to determine a mixture of multiple network structures.
Notice that most ABMs we reviewed so far either assume a single underlying social network (with parameters determined in model calibration) or generate artificial networks based on empirical findings or social science theories.  
They first choose a collection of feasible networks that represent the unobserved consumer networks. 
Then, a simple SIR model (similar to the Bass ABM in~\cite{herrmann13}) is used to simulate the diffusion of products.
The simulated time series are further transformed to multivariate stochastic functions, which provide priors to the Bayesian inference model to obtain the posterior weights on the set of feasible consumer networks.  
Like~\cite{herrmann13}, the adoption model is calibrated from the aggregate output, rather than from observations of individual decisions.

\citet{chica17} 
propose an agent-based framework to build decision support system (DSS) for word-of-mouth programs.
They developed a DSS to forecast the purchase of a freemium app and evaluate marketing policies, such as targeting and reward. 
The model captures seasonality of user activities by two probabilities for weekday and weekend respectively.
The initial social network is generated by matching the degree distribution of the real network. 
Then, for each node, two weights are assigned to in- and out-edges, respectively, turning the network into a weighted graph that represents the heterogeneous social influence among social neighbors. 
Specifically, two models are used to model the information diffusion. 
One is the Bass-ABM (~\cite{rand11}); the other is a contagion model (a threshold model but adding external influence). 
The parameters of the model were calibrated by a genetic algorithm~\citep{stonedahl14}, in which the fitness is defined based on the difference of simulated adoption from the historical adoption trajectory.
Notably, the model was validated by a hold-out dataset, which is independent of the training data. For example, the entire 3 month period spanned by the data was divided into two: first 60 days for training, the last 30 days for validation.

The independent cascade model used by \citet{herrmann13} and the threshold model used by~\citet{chica17} are significant insofar as these connect to a substantial literature that has recently emerged within the Computer Science community on \emph{information diffusion}, whereby information (broadly defined) spreads over a social network.
We make this connection more precisely in Section~\ref{sec:id} below.

\section{Categorization of Innovation Diffusion Models by Application}

Thus far, we followed a categorization of agent-based models of innovation diffusion focused on methods by which agent behavior is modeled.
First, we observe that methods range from sophisticated mathematical optimization models (Section~\ref{se: mo}), to economic models (Section~\ref{se: ec}), to even simpler models based on heuristics for representing agent behavior (Section~\ref{se:hr}). 
While economic factors are dominant concerns in some applications, others emphasize the cognitive aspects of human decision-making (Section~\ref{se:ca}) and are frequently used to model influence over online social networks (Section~\ref{se:si}).
Second, we note that the method chosen to capture agent behavior also impacts the techniques used to calibrate model parameters from data.
For example, cognitive models are often constructed based on detailed behavior data collected from field experiments and surveys, whereas models of agent behavior based on statistical principles rely on established statistical inference techniques for model calibration based on individual behavior data that is either observational or experimental. 
Other modeling approaches within our six broad categories often do not use data to calibrate individual agent behavior, opting instead to tune model parameters in order to match aggregate adoption data.

We now offer an alternative perspective to examine the literature on empirical ABMs of innovation diffusion by considering applications---that is, what particular innovation is being modeled.
A breakup of existing work using this dimension is given in Table~\ref{tb:app}.
As shown in the first column, we group applications by broad categories: agricultural innovations and farming, sustainable energy and conservation technologies, 
consumer technologies and innovations, information technologies and
social goods. 
Interestingly, the first two categories account for more than half of the publications in literature. 
This likely reflects the history of ABM as an interdisciplinary modeling framework for computational modeling of issues that are of great interest in social science.
A closely related factor could be the relatively high availability of data in these applications generated by social scientists (e.g., through the use of surveys).
Another interesting observation that arises is methodological convergence for given applications: relatively few applications have been modeled within different methodological frameworks as categorized above.
Future research may explore the use of different methods for same application. 
Furthermore, comparison of different modeling methods is rare within a single work (except in~\citep{dugundji08, zhang16}), although such a methodological cross-validation is of importance as emphasized by some authors~\citep{carley96, rand11}.   

\fontsize{6}{8}\selectfont
\begin{longtable}{|p{25mm}|l|p{30mm}|p{28mm}|}
\hline
\label{tb:app}
Category & Application & Method & Citation\\
\hline
agricultural innovations \newline and farming 
&agricultural innovations&mathematical programming&\citet{berger01, berger07, berger07wrm, schreinemachers09,berger10,alexander13}\\
&&economic (utility) & \citet{holtz12}\\
&organic farming& cognitive model (Deffuant) &\citet{deffuant02}\\
&&cognitive model (TPB, Deffuant) &\citet{kaufmann09}\\
&biogas plant&economic (profit)& \citet{sorda13}\\
&payments for ecosystem services & machine learning &  \citet{sun13}\\
\hline
sustainable energy and \newline conservation technologies
&water-saving technology & cognitive model (TPB) & \citet{schwarz09}\\
&&discrete choice model & \citet{galan09}\\
&heating system    &    cognitive model (TPB) 
& \citet{sopha13}\\
&&   conjoint analysis            & \citet{lee14}\\
&&   economic (cost)                & \citet{faber10}\\
&solar photovoltaic & heuristic    & \citet{zhao11}\\
&& economic (utility)        & \citet{palmer15}\\
&&cognitive model (TPB, Deffuant) & \citet{rai15}\\
&& machine learning & \citet{zhang16}\\            
&fuel cell vehicles & cognitive model (Consumat) & \citet{schwoon06}\\
&energy innovation & discrete choice model & \citet{tran12}\\
&electric vehicles  & cognitive model (TEC) & \citet{wolf12}\\
&&economic (utility) & \citet{plotz14}\\
&&economic (utility) & \citet{mccoy14}\\
&alternative fuel vehicles & conjoint analysis & \citet{zhang11}\\
&alternative fuels        & heuristic    &\citet{vliet10}\\
&&economic (utility)& \citet{gunther11}\\
&&conjoint analysis & \citet{stummer15}\\
&green electricity & cognitive model (LARA) &\citet{krebs15, ernst16}\\
&air-quality feedback device & cognitive model (TPB) & \citet{jensen16}\\
\hline
consumer technologies\newline and innovations 
&wine bottle closures & conjoint analysis & \citet{garcia07}\\
&mobile phones& conjoint analysis & \citet{vag07}\\
&transportation mode & discrete choice model & \citet{dugundji08}\\    
&new cars    & Fuzzy TOPSIS (heuristic) Model    & \citet{kim11}\\
&movie & economic (utility)    & \citet{broekhuizen11}\\
\hline
information technologies & Skype & social contagion model & \citet{karsai14, karsai16}\\
& Twitter & independent cascade model & \citet{herrmann13}\\
& Facebook app & social contagion model & \citet{trusov13}\\
& freemium app & social contagion model & \citet{chica17}\\
\hline
social goods & neighborhood support & cognitive model (LARA) & \citet{krebs13}\\ 
\hline
\caption{Categorization of surveyed work by Applications}\\
\end{longtable}
\normalsize

\section{Information Diffusion Models}
\label{sec:id}
Online social networks have emerged as an crucial medium of communication. 
It does not only allow users to produce, exchange, and consume information at an unprecedented scale and speed, but also speeds the diffusion of novel and diverse ideas~\citep{guille13, bakshy12}.
The emergence of online social networks and advances in data science and machine learning have nourished a new field: \emph{information diffusion}.
The fundamental problem in information diffusion is to model and predict how information is propagated through interpersonal connections over social networks using large-scale diffusion data. 
In fact, several authors have reviewed the topic of information diffusion over online social networks~\citep{bon11, guille13, shakarian15}. 
Our aim is not to provide a comprehensive review of this same topic.
Instead, we are interested in building connections between the agent-based modeling approach to innovation diffusion, and the modeling methods in the field of information diffusion. 
Indeed, researchers in the ABM community have paid little attention to the existing methods for modeling information diffusion, and especially in the played by data science in this field, which has significant implications for ABM model calibration, as we discuss below.

\subsection{Two Basic Models of Information Diffusion}
Compared to agent adoption models in Section~\ref{sec: emp_abms}, the decision process in the information diffusion literature is typically very simple, following predominantly the social influence models. 
The two most common models in information diffusion are Independent Cascades (IC)~\citep{goldenberg01} and Linear Threshold (LT) models~\citep{granovetter78}. 
These models are defined on directed graphs where activation is assumed to be monotonic: once a node is active (e.g., adopted, received information), it cannot become inactive.
The diffusion process in both models starts with a few active nodes and progresses iteratively in a \emph{discrete} and \emph{synchronous} manner until no new nodes can be infected. 
Specifically, in each iteration, a new active node in the IC model is given a single chance to activate its inactive neighbors independently with an exogenously specified probability (usually represented by the weight of the corresponding edge).
In the LT model, in contrast, an inactive node will become active only if the sum of weights of its activated neighbors exceeds a predefined node-specific threshold, which is typically randomly assigned between 0 and 1 for each network node. 
Note that in both models a newly activated node becomes active immediately in the next iteration.
From an agent-based perspective, both IC and LT are generative models which define two diffusion mechanisms.


\subsection{Learning Information Diffusion Models}
\label{sse:lidm}
Several efforts use empirical data to calibrate the parameters of the LT and IC models.
\citet{saito11} propose an asynchronous IC (AsIC) model, which not only captures temporal dynamics, but also node attributes. They show how the model parameters can be estimated from observed diffusion data using maximum likelihood estimation (MLE). 
The AsIC model closely follows the IC model, but additionally introduces a time delay before a newly activated node becomes active.
The time delay is assumed to be exponentially distributed with a parameter that is defined as an exponential function of a feature vector (a composition of attributes associated with both nodes and edges).  
The transmission probability is then defined as a logit function of the feature vector.
The data is given in the format of ``diffusion traces'', and each trace is a sequence of tuples which specify activation time for a subset of nodes.
To learn the model using this data, the authors define the log-likelihood of the data given the model. 
The authors then demonstrate how to solve the resulting optimization problem using expectation-maximization (EM).
While the proposed model is promising to be used for prediction, the learning method was only tested using synthetic data.

\citet{guille12} show how to parameterize the AsIC model using machine learning methods based on Twitter data.
In their model, the diffusion probability for information at any given time between two users is a function of attributes from three dimensions: \emph{social}, \emph{semantic}, and \emph{time}, which group features with respect to social network, content and temporal property respectively.
Four different classifiers were trained and compared in terms of cross-validation error: C4.5 decision tree, linear perceptron, multilayer perceptron, and Bayesian logistic regression.
The last model mentioned above was finally used for prediction.
Notably, time-delay parameter was determined separately in this work by comparing simulation results with actual diffusion dynamics, which is the same calibration method commonly used in ABM of innovation diffusion.
Unlike~\citep{saito11}, where all model parameters are inferred by MLE,  
here only a subset of model parameters are estimated through established machine learning techniques, but the rest are calibrated by simulations. 
Their evaluation shows that the model accurately predicts diffusion dynamics, but fails to accurately predict the volume of tweets. 
In our ABM jargon, the model performs well at macro-level, but poorly at micro-level validation~\citep{carley96}. 
Another limitation of this work is that validation is only performed as an in-sample exercise, rather than using out-of-sample data.

\citet{galuba10} propose two diffusion models with temporal features that are used to predict user re-tweeting behaviors on Twitter.
Both models define the probability for a user to re-tweet a given URL to be a product of two terms: one is \emph{time-independent}, the other is \emph{time-dependent}. 
Both have the same time-dependent part which follows a log-normal distribution, but differ in the actual definitions of the time-independent part.   
In their first model termed At-Least-One (ALO), the time-independent component is defined as the likelihood of at least one of the causes: either one is affected by the agent it follows, or by the user tweets a URL spontaenously.
The second, Linear Threshold (LT), model, posits that a user re-tweets a URL only if the cumulative influence from all the followees is greater than a threshold. The time-independent component in this model is given by a sigmoid function.
In order to calibrate and validate the model, the data set was split along the time dimension into two parts.
The model was calibrated 
by choosing parameters that optimize the estimated F-score using the gradient ascent method on the first (earlier) data set, and used to predict URL mentions in the second (later) data set.
Their results show that the LT model achieves the highest F-score among all models and correctly predicts approximately half of URL mentions 
with lower than 15\% false positives. 

While all research reviewed so far assumes known network structure, a number of efforts deal with hidden network structures which must also be learned from data.
The so-called \emph{network inference problem} is to infer the underlying network given a complete activation sequence~\citep{guille13}. 
\citet{gomez10} introduce a variant of the independent cascade model~\citep{kemp03} adding time delay. 
Their problem is to find a directed graph with at most $k$ edges that maximizes the likelihood of a set of cascades for a given transmission probability and parameters of the incubation distribution, which is solved approximately using a greedy algorithm.
\citet{myer10} propose a cascade model which is similar to~\citet{gomez10} but allows distinct transmission probabilities for different network edges.
The goal is to infer the adjacency matrix (referring to the pairwise transmission probabilities) that maximizes the likelihood given a set of cascades, which is accomplished by solving a convex optimization problem derived from the problem formulation.
\citet{gomez11} develop a continuous-time diffusion model that unifies the two-step diffusion process involving both a transmission probability and time delay from~\citet{gomez10} and~\citet{myer10}. 
The pivotal value is the \emph{conditional} probability for a node $i$ to be infected at time $t_i$ given that a neighboring node $j$ was infected at time $t_j$, which is formulated as a function of the time interval $(t_i-t_j)$ and parametrized by a pairwise transmission rate $\alpha_{ji}$.
Survival analysis~\citep{lawless11}~is used to derive the maximum likelihood function given a set of cascades, and they aim to find a configuration of all transmission rates that maximizes the likelihood.
While most network inference algorithms assume static diffusion networks,~\citet{gomez13} address a network inference problem with a time-varying network.
The resulting inference problem is solved using an online algorithm upon formulating the problem as a stochastic convex optimization. 
\leaveout{
Although most network inference algorithms assume static diffusion networks,  
\citet{gomez13} develop an algorithm termed \emph{InfoPath} for time-varying network inference
to model information diffusion in online social networks.
The method generalizes the static network inference approach by~\citet{gomez11} to dynamic networks with edge transmission rate $\alpha_{j,i}(t)$ that may change over time. 
The \emph{dynamic network inference problem} is solved by an online algorithm based on a formulation of stochastic convex optimization. 
}

\subsection{Bridging Information Diffusion Models and Agent-Based Modeling of Innovation Diffusion}
The methodological framework of the information diffusion inference problems discussed above is a natural fit for principled data-driven agent-based modeling.
The information diffusion models characterized by transmission probabilities and time delay are essentially agent-based models. 
Given data of diffusion cascades, they can be constructed either using only the temporal event (adoption) sequence, or using more general node features, social network, content, and any other explanatory or predictive factors. 
In fact, ABM researchers have started to apply similar statistical methods to develop empirical models (see Section~\ref{se: st}).
Notably, as shown by~\citet{zhang16}, parametric probabilistic models of agent behavior can be estimated from observation data using maximum likelihood estimation methods.
In addition, the approaches for network inference appear particularly promising in estimating not only behavior for a known, fixed social influence network, but for estimating the influence network itself, as well as the potentially heterogeneous influence characteristics.

A crucial challenge in translating techniques from information diffusion domains to innovation diffusion is that in the latter only observes a single, partial adoption sequence, rather than a collection of complete adoption sequences over a specified time interval.
As a consequence, the fully heterogeneous agent models cannot be inferred, although the likelihood maximization can still be effectively formulated by limiting the extent of agent heterogeneity (with the limit of homogeneous agents used by~\citet{zhang16}).
In addition, the assumptions generally made in information diffusion models can also pose serious challenges to the transferability of the approach to agent-based modeling. 
Recall that the information cascade models often assume that an adopter has a \emph{single} chance to affect its inactive neighbors and a non-adopter is affected by its neighboring adopters \emph{independently}. 
These assumptions simplify the construction of the likelihood function, but further justification is needed for them, especially when building empirical models that are expected to faithfully represent realistic social systems and diffusion processes. 
Note that rules that govern the interactions in agent-based models are quite flexible and can be very sophisticated, which is also one of the major advantages of agent-based computing over analytical models.  
Although one may be able to explicitly derive a parametric likelihood function given diffusion traces in more complex settings than existing information diffusion models do, this is sure to be technically challenging. 
Moreover, solving the resulting MLE can be computationally intractable.    
Therefore, to take advantage of MLE approach in information diffusion, ABM researchers must make appropriate assumptions on agent interactions so that they can derive tractable likelihood functions without significantly weakening the model's explanatory and predictive power.

\section{Discussion}


\subsection{Validation in Agent-Based Modeling}
\label{sse:cv}

As agent-based modeling is increasingly called for in service of decision support and prediction, it is natural to expect them to be empirically grounded.
An overarching consideration in empirically grounded agent-based modeling is how data can be used in order to develop reliable models, where reliability is commonly identified with their ability to accurately represent or predict the environment being modeled.
This property of \emph{reliability} is commonly confirmed through model \emph{validation}.
In social science, a number of authors have contributed to the topic of validation, from approaches for general computational models~\citep{carley96}, to those focused on agent-based simulations~\citep{xiang05, fagiolo06, garcia07, ormerod09, rand11}, to specific types of agent-based models~\citep{brown05}. 
Outside of social science, validation of simulation systems has an even longer history of investigation~\citep{knepell93, banks98, kleijen99, sanchez01}.
We now briefly review these approaches.

As previously mentioned, \citet{carley96} suggests four levels of validation: \emph{grounding}, \emph{calibration}, \emph{verification}, and \emph{harmonizing}.
\emph{Grounding} establishes reasonableness of a computational model, including face validity, parameter validity, and process validity; 
\emph{calibration} establishes model's feasibility by tuning a model to fit empirical data; 
\emph{verification} demonstrates how well a model's predictions match data; and \emph{harmonization} examines the theoretical adequacy of a verified computational model.

More recently, drawing on formal model verification and validation techniques from industrial and system engineering for discrete-event system simulations, \citet{xiang05} suggest the software implementation of agent-based model has to be verified with respect to its conceptual model, and highlight a selection of validation techniques from~\citet{banks98}, such as face validation, internal validation, historical data validation, parameter variability, predictive validation, and Turing tests. Moreover, they suggest the use of other complementary techniques, such as model-to-model comparison~\citep{axtell96} and statistical tests~\citep{kleijen99, sanchez01}. 

For agent-based models in economics, \citet{fagiolo06} proposed three different types of \emph{calibration} methods: the \emph{indirect calibration} approach, the \emph{Werker-Brenner empirical calibration} approach, and the \emph{history-friendly} approach. 
For example, \citet{garcia07} adopt the last approach to an innovation diffusion study in New Zealand winery industry, using conjoint analysis to instantiate, calibrate, and verify the agent-based model qualitatively using stylized facts.

For agent-based models in marketing, \citet{rand11} suggest verification and validation as two key processes as guidelines for rigorous agent-based modeling. 
The use of term ``verification'' follows common understanding in system engineering~\citep{xiang05}. 
In particular, the authors identify four steps for validation: micro-face validation, macro-face validation, empirical input validation, and empirical output validation using stylized facts, real-world data, and cross-validation. 
Note that the proposed validation steps echo the framework by~\citet{carley96}: the first two steps correspond to grounding, the third to calibration, and the fourth roughly combines verification and harmonization. 
However, the cross-validation method mentioned in~\citet{rand11} appears to suggest validation across models, whereas \citet{carley96} suggests validation across multiple data sets. 
The latter is consistent with the use of cross-validation in statistical inference and machine learning~\citep{friedman01, bishop06}. 

Focusing specifically on empirically grounded ABMs, we suggest two pivotal steps in ensuring model reliability in a statistical sense: \emph{calibration} and \emph{validation}.
By calibration, we mean the process of \emph{quantitatively} fitting a set of model parameters to data, whereas validation means a quantitative assessment of the predictive efficacy of the model \emph{using independent data}, that is, using data which was not utilized during the calibration step.
Moreover, insofar as a model of innovation diffusion is concerned with predicting future diffusion of an innovation, we propose to further split the dataset along a temporal dimension, so that earlier data is used exclusively for model calibration, while later data exclusively for validation.
Starting with this methodological grounding, we now proceed to identify common issues that arise in prior research on empirically grounded agent-based models of innovation diffusion.

\subsection{Issues in Model Calibration and Validation}
\label{sse: is}

Agent-based modeling research has often been criticized for lack of accepted methodological standard, hindering its acceptance in top journals by mainstream social scientists. One notable protocol due to~\citet{richiardi06} highlight four potential methodological pitfalls: \emph{link with the literature}, \emph{structure of the models}, \emph{analysis}, and \emph{replicability}. 

A careful examination of the empirical ABM work on innovation diffusion through this protocol suggests that most of these issues have been addressed or significantly mitigated.
For example, nearly all of the reviewed papers present theoretical background, related work, sufficient description of model structure, sensitivity analysis of parameter variability, a formal representation (e.g., UML\footnote{The short for the Unified Modeling Language, developed by the Object Management Group: http://www.omg.org}, 
OOD\footnote{A standard to describe agent-based models originally proposed by~\citet{grimm06} 
for ecological modeling.}), and public access to source code. 
In spite of these improvements, however, there are residual concerns about systematic quantitative calibration and validation using empirical data.

We observe that different agent adoption models are calibrated differently. In the case of cognitive agent models (Section~\ref{se:ca}), such as the Theory of Planned Behavior and theory of emotional coherence, the individual model parameters are often estimated using survey data.
Similarly, statistics-based models (Section~\ref{se: st}) can be parametrized using either experimental or observational individual-level data. 
On the other hand, for conceptual models, such as heuristic (Section~\ref{se:hr}) and economic models (Section~\ref{se: ec}), calibration is commonly done by iteratively adjusting parameters to match simulated diffusion trajectory to aggregate-level empirical data. 
Formally, we call the first kind of calibration ``\emph{micro-calibration}'', as it uses individual data during calibration, whereas the second type ``\emph{macro-calibration}'', as it uses aggregate-level data. 
Moreover, in many studies simulation parameters are determined using both micro- and macro-calibration. For example, since network structure is often not fully observed, and rules that govern agent interactions are assumed, parameters of these are commonly macro-calibrated. 
Our first concern is about macro-calibration. 

\noindent{\bf Issue I: Potential pitfalls in macro-calibration.} 
When a model has many parameters, over-fitting the model to data becomes a major concern~\cite{friedman01, bishop06}.
As~\citet{carley96} suggests, ``any model with sufficient parameters can always be adjusted so that some combination of parameters generates the observed data, therefore, large multi-parameter models often run the risk of having so many parameters that there is no guarantee that the model is doing anything more than curve fitting.''
Interestingly, the issue of over-fitting may even be a concern in macro-calibration when only a few parameters need to be calibrated.
The reason is that agent-based models are highly non-linear, and even small changes in several parameters can give rise to substantially different model dynamics.
This issue is further exacerbated by the fact that macro-calibration makes use of aggregate-level data, which is often insufficient in scale for reliable calibration of any but the simplest models, as many parameter variations can give rise to similar aggregate dynamics.

Addressing the issue requires greater care and rigor in applying macro-calibration.
One possibility is that instead of choosing only a single parameter configuration, to select a parameter \emph{zone} using a classifier such as decision trees~\citep{kaufmann09} or other machine learning algorithms. 
Subsequently, the variability of parameters within this zone can be further investigated using sensitive analysis. 
Another potential remedy is that instead of using only a single target statistic (e.g., average adoption rates) to use multiple indicators. 
A relevant strategy to build agent-based models in the field of ecology is termed ``pattern-oriented modeling'', which utilizes multiple patterns at different scales and hierarchical levels observed from real systems to determine the model structure and parameters~\citep{grimm05}.


In addition, there are more advanced and robust techniques that can improve the rigor of macro-calibration. 
The modeling framework in~\citep{zhang16} and statistical inference methods introduced in Section~\ref{sec:id} propose methods which integrate micro and macro calibration into a single maximum likelihood estimation framework. Through well-established methods in machine learning, such as cross-validation, one can expect to parameterize a highly-predictive agent-based model and minimize the risk of over-fitting.
Indeed, a fundamental feature of any approach should be to let validation ascertain the effectiveness of macro-calibration in generalizing beyond the calibration dataset.
This brings us to the second common issue revealed by our review: lack of validation on independent data.


\noindent{\bf Issue II: Rigorous quantitative validation on independent data is uncommon.}
A common issue in the research we reviewed is that validation is often informal, incomplete, and even missing.
The common reason for incomplete data-driven validation is that relevant data is simply unavailable.
However, so long as data is available for calibrating the model, one can in principle use this data for both calibration and validation steps, for example, following cross-validation methods commonly utilized in machine learning.
Several efforts seek to standardize the validation process for agent-based models, and computational models in general. 
However, few papers discussed explicitly follow any formalized validation approaches in this literature, although important exceptions exist~\citep{garcia07, zhang11, stummer15}. 


\noindent{\bf Issue III: Few conduct validation at both micro-level and macro-level.} 
There has been some debate about whether validation should be performed at both micro- and macro-level~\citep{carley96}. 
While arguments against the dual-verification often emphasize greater importance of model accuracy at the aggregate level, we argue that robust predictions at the aggregate level can only emerge when individual behavior is accurately modeled as well, particularly when policies that the ABM evaluates can be implemented as modifying individual decisions.

Statistics-based models, such as machine learning, have well-established validation techniques which can be leveraged to validate individual-level models. 
One widely-used technique in machine learning and data mining is \emph{cross-validation}. 
A common use of cross-validation is by partitioning the data into $k$ parts, with training performed on $k-1$ of these and testing (evaluation) on the $k$th.
The results are then averaged over $k$ independent runs using each of the parts as test data.
Observe that such a cross-validation approach can be used for models of individual behavior that are not themselves statistically-driven, such as models based on the theory of planned behaviors. 
Unfortunately, few of the surveyed papers, with the exception of statistics-based models, use cross-validation. 

\noindent{\bf Issue IV: Few conduct validation of forecasting effectiveness on independent ``future'' data.}
One limitation of cross-validation techniques as traditionally used is that they provide an offline assessment of model effectiveness.
To assess the predictive power of dynamical systems, the entire model has to be validated in terms of its ability to predict ``future'' data relative to what was used in calibration. 
We call this notion ``\emph{forward validation}''. 
In particular, forward validation must assess simulated behaviors against empirical observations at both individual and aggregate levels \emph{with an independent set of empirical data}.
This can be attained, for example, by splitting a time-stamped data set so that calibration is performed on data prior to a split date, and forward validation is done on data after the split date~\citep{galan09, zhang16, rai15, chica17}.
In this review, we do observe several approaches that are validated on independent data, but these either are not looking forward in time relative to the calibration data, or only focus on macro-level validation.
A common argument for the use of in-sample data for the forward validation is that new data is not available while the modeling task is undertaken.
Notice, however, that any data set that spans a sufficiently long period of time can be split along the time dimension as above to effect rigorous forward validation.

\subsection{Recommended Techniques for Model Calibration and Validation}
We have identified several issues in calibration and validation which commonly arise in the prior development of empirical agent-based models for innovation diffusion, and briefly discussed possible techniques that can help address these issues. 
We now summarize our recommendations:
\begin{enumerate}
\item[] \emph{Multi-Indicator Calibration}. When macro-calibration is needed, the use of multiple indicators can help address \emph{over-fitting}, whereby a model which appears to effectively \emph{match} data in calibration performs poorly in prediction on unseen data.
We suggest that such indicators are developed at different scale and hierarchical levels, so that models which cannot effectively generalize to unseen data can be efficiently eliminated.
\item[] \emph{Maximum Likelihood Estimation}. When individual-level data are available, we recommend constructing probabilistic adoption models for agents, and estimating parameters of these models by maximizing a global likelihood function (see, for example, the modeling framework by~\citet{zhang16}, and research discussed in Section~\ref{sse:lidm}). Doing so offers a principled means of calibrating agent behavior models from empirical data.
\item[]    \emph{Cross Validation}. This approach is widely used for model selection in the machine learning literature. Here, we recommend it for both micro-calibration and micro-validation of ABMs. Note that it does not only apply to statistics-based models, but can be used for any agent modeling paradigm where model parameters are calibrated using empirical data. The use of cross-validation in calibration can dramatically reduce the risk of over-fitting. Moreover, as it inherently uses independent data, such validation leads to more rigorous ABM methodology.
\item[]    \emph{Forward Validation}. This method involves splitting data into two consecutive time periods. The modeler calibrates an agent-based model using data from the first period, and assesses the predictive efficacy of the model in the second period. More rigorously, validation of the model should be evaluated at both individual and aggregate levels. 
\end{enumerate}

\section{Conclusions}
We provided a systematic, comprehensive, and critical review of existing work on empirically grounded agent-based models for innovation diffusion. 
We offered a unique methodological survey of literature by
categorizing agent adoption models along two dimensions: methodology
and application.
We identified six methodological categories: \emph{mathematical optimization based models}, \emph{economic models}, \emph{cognitive agent models}, \emph{heuristic models}, \emph{statistics-based models} and \emph{social influence models}. 
They differ not only in terms of assumptions and elaborations of human
decision-making process, but also with respect to calibration and
parameterization techniques. 
Our critical assessment of each work focused on using data for calibration and
validation, and particularly performing validation with independent data. 
We briefly reviewed the most important work in the closely related
literature on information diffusion, building connections between the
innovation and information diffusion approaches. 
One particularly significant observation is that information diffusion
methods rely heavily on machine learning and maximum likelihood
estimation approaches, and the specific methodology used can be
naturally ported to innovation diffusion ABMs.
Drawing on prior work in validation of computational models, we discussed four main issues for existing  empirically grounded ABM studies in innovation diffusion, and provided corresponding solutions.

On balance, recent developments of empirical approaches in agent-based
modeling for innovation diffusion are encouraging. 
Although calibration and validation issues remain in many studies, a
number of natural solutions from data analytics offer promising
directions in this regard.   
The ultimate goal of empirically grounded ABMs is to provide decision
support for policy makers and stakeholders across a broad variety of
innovations, helping improve targeted marketing strategies, and reduce
costs of successful translation of high-impact innovative technologies
to the marketplace.

\section{Acknowledgments}
This  work  was  partially  supported  by  the  U.S.  Department of Energy (DOE) office of Energy Efficiency and Renewable Energy, under the Solar Energy Evolution and Diffusion Studies (SEEDS) program, the National Science Foundation (IIS-1526860), and the Office of Naval Research (N00014-15-1-2621).

\bibliographystyle{spbasic}      
\bibliography{egabm}   

%
%

\end{document}